\documentclass[%
 reprint,
 amsmath,amssymb,
 aps,
]{revtex4-2}

\usepackage{graphicx}% Include figure files
\usepackage{dcolumn}% Align table columns on decimal point
\usepackage{bm}% bold math
\usepackage{amsmath,amssymb,amsfonts,amsthm,bm,bbm,cancel,wasysym}
\usepackage{epsfig,graphics,graphicx,epstopdf,caption,subcaption}
\graphicspath{{Charts/}}
\usepackage{array,booktabs,colortbl,colordvi,multirow}
\usepackage{colordvi,color,xcolor}
\usepackage{hyperref}
\usepackage{rotating}
\usepackage{comment}
\usepackage{feynmp-auto}
\usepackage{verbatim}
\usepackage{setspace}
\usepackage{url}
\usepackage[percent]{overpic}
\usepackage{slashed}
\usepackage{xspace}
\usepackage{fullpage}
\newskip\zatskip \zatskip=0pt plus0pt minus0pt
\def\matth{\mathsurround=0pt}

\def\gsim{\mathrel{\mathpalette\atversim>}}
\def\atversim#1#2{\lower0.7ex\vbox{\baselineskip\zatskip\lineskip\zatskip
  \lineskiplimit 0pt\ialign{$\matth#1\hfil##\hfil$\crcr#2\crcr\sim\crcr}}}

\newif\ifdiagrams
\diagramstrue
\ifdiagrams

\else
  \excludecomment{fmffile}
\fi

\usepackage{hyperref} %Automatically links \label and \ref commands; Always load last
\hypersetup{
    colorlinks=true,       % false: boxed links; true: colored links
    linkcolor=red,          % color of internal links
    citecolor=blue,        % color of links to bibliography
    filecolor=magenta,      % color of file links
    urlcolor=blue           % color of external links
}
\usepackage[all]{hypcap} %Link navagates to top of figure instead of caption (below fig)
\begin{document}

\preprint{APS/123-QED}

\title{Portal Matter, Kinetic Mixing, and Muon $g-2$}

\author{George N. Wojcik}
\email{gwojcik@wisc.edu}
\author{Lisa L.~Everett}%
\email{leverett@wisc.edu}
\author{Shu Tian Eu}
\email{eu@wisc.edu}
\author{Ricardo Ximenes}
\email{dossantosxim@wisc.edu}
\affiliation{%
Department of Physics, University of Wisconsin-Madison, Madison, WI 53706, USA\\
}%

\date{\today}% It is always \today, today,

\begin{abstract}
We present a minimal construction using leptonic portal matter that addresses the muon $g-2$ anomaly. While the chiral enhancement mechanism is reminiscent of that of fermiophobic $Z'$ gauge models, the parameter space motivated by the kinetic mixing/vector portal dark matter model paradigm is vastly different and can be readily explored in current and forthcoming experiments.  
%We made a model that lets portal matter deal with $g-2$! \GW{This is a placeholder}
\end{abstract}

\maketitle

%\tableofcontents

\section{\label{sec:intro}Introduction}

The anomalous magnetic moment of the muon $a_\mu \equiv (g-2)_\mu/2$ has long been known to present an intriguing possible probe of  physics beyond the Standard Model (SM). The first results of the FNAL measurement \cite{Muong-2:2021ojo} combined with %the results of 
the BNL measurement \cite{Muong-2:2006rrc}, now place the discrepancy between the measured value of $a_\mu$ and the theoretical expectation of the Standard Model \cite{Aoyama:2020ynm,Aoyama:2012wk,Aoyama:2019ryr,Czarnecki:2002nt,Gnendiger:2013pva,Davier:2017zfy,Keshavarzi:2018mgv,Colangelo:2018mtw,Hoferichter:2019mqg,Davier:2019can,Keshavarzi:2019abf,Kurz:2014wya,Melnikov:2003xd,Masjuan:2017tvw,Colangelo:2017fiz,Hoferichter:2018kwz,Gerardin:2019vio,Bijnens:2019ghy,Colangelo:2019uex,Blum:2019ugy,Colangelo:2014qya} at
\begin{align}\label{eq:delta-a}
    \Delta a_\mu = a_\mu^{\textrm{exp}}-a_\mu^{\textrm{SM}} =  (251 \pm 59) \times 10^{-11},
\end{align}
indicating a significance of $4.2\sigma$. Assuming this discrepancy is indeed due to new physics \footnote{Recent lattice calculations of the hadronic vacuum polarization contribution to $g-2$ \cite{Borsanyi:2020mff,Lehner:2020crt,Ce:2022kxy,Alexandrou:2022amy,Colangelo:2022vok} suggest that the discrepancy can be attributed to this contribution, at the price of tension with the data-driven HVP calculation \cite{Crivellin:2020zul,Keshavarzi:2020bfy,Colangelo:2020lcg}. As the tension between the data-driven and lattice HVP results has yet to be explained, we find it reasonable to consider the possibility of new physics providing the dominant source of the discrepancy.}, many models have been proposed to address it \footnote{For reviews, see e.g. \cite{Jegerlehner_2009} and the more recent \cite{Athron:2021iuf}.  Models that were of particular notice for us in this work include ~\cite{deGiorgi:2022xhr,CarcamoHernandez:2019ydc,Dermisek:2013gta,Bodas:2021fsy,Agrawal:2022wjm,Endo:2021zal,Arcadi:2021cwg}.}.

One class of models that has received limited attention in this context 
%the context of the $g-2$ anomaly 
is the so-called ``portal matter'' models \cite{Rizzo:2018vlb,Holdom:1985ag,Holdom:1986eq,Wojcik:2020wgm,Wojcik:2021xki,Rueter:2019wdf,Rueter:2020qhf,Rizzo:2022lpm,Rizzo:2022qan}. Portal matter models address the $O(10^{-(3-5)})$ kinetic mixing required between the SM hypercharge and a hidden $U(1)_D$ symmetry, required in vector portal/kinetic mixing realizations of dark matter \cite{Holdom:1985ag,Holdom:1986eq,Pospelov:2007mp,Izaguirre:2015yja,Essig:2013lka,Curtin:2014cca}, by positing that kinetic mixing is generated at one loop by heavy new fields that are charged under both SM hypercharge and $U(1)_D$ (the portal matter fields) 
%via a vacuum polarization diagram 
as in \cite{Holdom:1985ag,Holdom:1986eq}. As noted in \cite{Rizzo:2018vlb}, if these fields are fermions that are light enough to be produced at collider scales, several arguments suggest that %if such fields are fermionic, 
they should be vector-like with the same SM quantum numbers as some SM fields, but with non-zero $U(1)_D$ charges. It was found that a minimal realization of portal matter is incapable of resolving the observed $g-2$ correction \cite{Rizzo:2018vlb}. 

However, the field content of the portal matter construction, specifically the addition of vector-like fermions charged under a hidden $U(1)_D$, bears a striking similarity to models which address the muon $g-2$ in a well-studied fermiophobic $Z'$ scenario \cite{King:2017anf,CarcamoHernandez:2019ydc,Lee:2022sic,Lee:2022nqz}, in which a chirally enhanced contribution to $\Delta a_\mu$ is obtained by adding both isospin doublet and isospin singlet vector-like leptons, charged under the $Z'$ gauge group, which can mix via a Yukawa coupling to the SM Higgs. In contrast to models with electroweak/TeV-scale $Z'$ bosons, the dark matter phenomenology in portal matter models strongly motivates a \emph{sub}-GeV dark gauge boson, and to our knowledge models with such a chirally enhanced $\Delta a_\mu$ value have not been explored in this region of parameter space. In this letter, we consider the correction to $\Delta a_\mu$ in a simple portal matter construction with the same chiral enhancement mechanism employed in e.g.~\cite{CarcamoHernandez:2019ydc}. We will see that the observed value of $\Delta a_\mu$ can easily be accomodated and, crucially, that this scenario can trivially satisfy existing phenomenological constraints, even those that present non-trivial challenges to the analogous constructions with heavier new gauge bosons.

\section{\label{sec:model-setup} Model Setup}
The model considered here is a simple extension of the minimal portal matter construction of \cite{Rizzo:2018vlb}. The SM gauge group is extended by an Abelian dark gauge symmetry $U(1)_D$ with an $O(1)$ gauge coupling constant $g_D$. The SM fermion content is neutral under $U(1)_D$, but couplings to the gauge boson of the $U(1)_D$, which we shall henceforth refer to as the dark photon $A_D$, and the SM fields are achieved by a small kinetic mixing coefficient $\epsilon \sim O(10^{-(3-5)})$, which to an excellent approximation imparts all SM fields with a coupling to the dark photon of $\epsilon$ times their photon coupling. The dark sector consists of the dark photon $A_D$, a SM singlet complex scalar $S$ with a charge of $+1$ under $U(1)_D$ that achieves a vacuum expectation value (vev) and gives the dark photon a mass, and a dark matter field $\chi$, which is a SM singlet charged under $U(1)_D$.

It is well known \cite{Izaguirre:2015yja,Darme:2017glc,Rizzo:2018ntg} that simplified dark matter models can generate the appropriate relic abundance consistent with other experimental constraints via $s$-channel annihilation of dark matter pairs into SM fermions, if the dark photon and dark matter both have masses of $O(0.01-1 \; \textrm{GeV})$ and $g_D \epsilon \sim O(10^{-4})$. As %we shall find that 
the phenomenology of interest here is largely agnostic to the parameters governing the dark matter relic abundance (provided they lie broadly within the range we have discussed), we shall not specify the spin or mass of the dark matter field and only assume it is selected so that the required relic abundance is reproduced \footnote{See e.g.~\cite{Yang:2022zlh,Amiri:2022cbv} for alternative scenarios in which the sub-GeV vector boson is the dark matter; however, in these scenarios kinetic mixing does not play an important phenomenological role.}.

Our primary focus in this work is the heavy portal matter fermions that generate kinetic mixing between $U(1)_D$ and the SM hypercharge group $U(1)_Y$ at one loop, as in \cite{Holdom:1985ag,Holdom:1986eq}. In this model, these consist of two new families of vector-like leptons -- two isospin doublet leptons $D^+$ and $D^-$, with $U(1)_D$ charges of $+1$ and $-1$, respectively, and two isospin singlet leptons $E^+$ and $E^-$, again with their $U(1)_D$ charges denoted by their superscripts. The detailed charge assignments are given in Table \ref{tab:fermion-content} \footnote{For simplicity, we do not include any SM singlet right-handed neutrinos. A comprehensive study of options for generating neutrino masses and lepton flavor mixing angles in this context is deferred to future work.}.

\begin{table}[]
    \centering
    \begin{tabular}{| c | c | c | c |}
        \hline
        Field & $SU(2)_L$ & $Y/2$ & $Q_D$ \\
        \hline
        $l_L = ( \nu^\mu_L, \mu_L)^T$ & $\mathbf{2}$ & $-\frac{1}{2}$ & $0$\\
        \hline
        $\mu_R$ & $\mathbf{1}$ & $-1$ & $0$\\
        \hline
        $L^+_{L,R} = ( \nu^+_{L,R}, D^+_{L,R})^T$ & $\mathbf{2}$ & $-\frac{1}{2}$ & $+1$\\
        \hline
        $E^+_{L,R}$ & $\mathbf{1}$ & $-1$ & $+1$\\
        \hline
        $L^-_{L,R} = (\nu^-_{L,R}, D^-_{L,R})^T$ & $\mathbf{2}$ & $-\frac{1}{2}$ & $-1$\\
        \hline
        $E^-_{L,R}$ & $\mathbf{1}$ & $-1$ & $-1$\\
        \hline
    \end{tabular}
    \caption{\footnotesize The representations of the portal matter leptons $D^\pm$ and $E^\pm$, and the second-generation SM leptons $l_L$ and $\mu_R$, with respect to $SU(2)_L \times U(1)_Y\times U(1)_D$. Note that all new fermions are vector-like under all gauge groups.}
    \label{tab:fermion-content}
\end{table}
For the set of vector-like fermion fields of SM hypercharges $Q_{Y_i}$, $U(1)_D$ charges $Q_{D_i}$, and masses $m_i$, the kinetic mixing coefficient $\epsilon$ is given by
\begin{align}\label{eq:KM-formula}
    \epsilon = c_W \frac{g_D g_Y}{12 \pi^2} \sum_i Q_{Y_i} Q_{D_i} \log \bigg( \frac{m_i^2}{\mu^2} \bigg),
\end{align}
where $\mu$ is the renormalization scale.
%
%We immediately see that the kinetic mixing of Eq.~(\ref{eq:KM-formula}) trivially guarantees that the new field content of Table \ref{tab:fermion-content} produces a finite and calculable kinetic mixing, as we would suspect if $U(1)_Y$ and $U(1)_D$ were part of a single unified group. Furthermore, for $g_D \sim O(1)$, the kinetic mixing is of the appropriate magnitude, $O(10^{-(3-5)})$, with the specific value of $\epsilon$ proportional to the dark gauge coupling $g_D$ and logarithmically dependent on the mass non-degeneracy between the $+1$ and $-1$ portal matter fields. Of course, there are many more complicated constructions which might generate a finite and calculable $\epsilon$ \cite{Rueter:2019wdf,Rizzo:2022lpm,Wojcik:2020wgm}, however for simplicity we adhere to this framework, reminiscent of the minimal setup of \cite{Rizzo:2018vlb}.

The Lagrangian of this model is given by 
%\begin{equation}
$\mathcal{L}=\mathcal{L}_G+\mathcal{L}_H+\mathcal{L}_f+\mathcal{L}_Y,$
%\end{equation}
for the gauge, Higgs, fermion, and Yukawa sectors, respectively. $ \mathcal{L}_{G}$ denotes the SM gauge Lagrangian together with the dark $U(1)$ contribution, as discussed.  
%%The gauge part of the Lagrangian can be expressed as
%\begin{equation}
%    \mathcal{L}_{G}=\mathcal{L}^{\text{SM}}_{G}-\frac{1}{4}F_{D\mu\nu}F_{D}^{\mu\nu},
%\end{equation}
%in which $F_D$ is the field strength tensor of the $U(1)_D$ gauge boson $A_D$. % and $\mathcal{L}^{\text{SM}}$ is the standard SM gauge Lagrangian. 
$\mathcal{L}_H$ includes the SM Higgs $H$ and dark Higgs Lagrangian terms, which are %The Higgs part of the Lagrangian is given by
\begin{equation}
    \mathcal{L}_{H}=\mathcal{L}^{\text{SM}}_{H}+(D_\mu S)^\dagger (D^\mu S)-V(H,S),
\end{equation}
%which includes the standard SM Higgs Lagrangian $\mathcal{L}^{\text{SM}}_{H}$ and the scalar potential
in which $V(H,S)$ takes the form
\begin{equation}
    V(H,S)=-\mu^2_S S^\dagger S+\lambda_S (S^\dagger S)^2+\lambda_{HS}(H^\dagger H)(S^\dagger S).
\end{equation}
Here we will assume $\lambda_{HS}\ll \lambda_{S},\lambda_{H}$ for simplicity \footnote{
 In this class of models, the mixing between the SM Higgs $H$ and a dark Higgs %, mediated by terms such as $\lambda H^\dagger H S^\dagger S$, 
must be suppressed (e.g.~$\lambda_{HS} \sim O(10^{-4})$) \cite{ATLAS:2019cid,Rizzo:2018vlb,Rizzo:2020ybl,Yang:2022zlh} to avoid overly large corrections to the invisible branching fraction of the SM Higgs or constraints on light hidden scalar searches \cite{Winkler:2018qyg}. Thus we have assumed for simplicity that there is no mixing between the dark Higgs and SM Higgs sector. 
Including a small but nonzero $\lambda_{HS}$ within allowed constraints does not appreciably change the phenomenology of the portal matter sector.}.
%, we will neglect $\lambda_{HS}$ for simplicity. 
The fermionic kinetic terms of $\mathcal{L}_f$ are %The fermionic terms of the Lagrangian is given by
\begin{equation}
    \mathcal{L}_{f}=\mathcal{L}^{\text{SM}}_{f}+\sum_{F=L^{\pm},E^{\pm}} i( \overline{F}_{L} \slashed{D}F_L+\overline{F}_R \slashed{D}F_R ).
\end{equation}
The Yukawa sector $\mathcal{L}_Y$ governs the mixing of portal matter fields $D^\pm$ and $E^\pm$ with the SM leptons.
For simplicity, we assume here that all portal matter fields only mix with the second-generation leptons $\mu$ and $\nu_\mu$, which may be achieved via some flavor symmetry. Relaxing this assumption and allowing portal matter couplings to e.g.~electrons, would introduce significant additional constraints on the model from lepton flavor-violating processes, similar to those discussed in \cite{CarcamoHernandez:2019ydc}. %The Yukawa sector of the second-generation SM leptons and the new portal matter fields, which governs the mixing between the SM and portal matter fields,  is then given by
Hence, $\mathcal{L}_Y$ is given by

\begin{widetext}
\begin{align}\label{eq:fermion-yukawa}
    \mathcal{L}_{\textrm{Y}} &  \supset  -y_{\mu} \overline{l}_L H \mu_R - y^+_{L} \overline{l}_L S^\dagger L^+_R - y^-_{L} \overline{l}_L S L^-_R - y^+_E \overline{E}^+_L S \mu_R - y^-_E \overline{E}^-_L S^\dagger \mu_R \nonumber\\
    &- y^+_{LE} \overline{L}^+_L H E^+_R - y^+_{EL} \overline{E}^+_L H^\dagger L^+_R - y^-_{LE} \overline{L}^-_L H E^-_R - y^-_{EL} \overline{E}^-_L H^\dagger L^-_R \\
    &- M^+_L \overline{L}^+_L L^+_R - M^+_E \overline{E}^+_L E^+_R - M^-_L \overline{L}^-_L L^-_R - M^-_E \overline{E}^-_L E^-_R + h.c., \nonumber
\end{align}
\end{widetext}
where the $y$'s are Yukawa couplings 
%, $H$ is the SM Higgs \RX{The Higgs was previously defefined, should we remove this line?}, 
and $M^\pm_{L,E}$ are vector-like mass parameters. Without loss of generality, we can perform phase rotations on the fields in Eq.~(\ref{eq:fermion-yukawa}) to guarantee that all $M$ parameters, as well as $y^\pm_{L,E}$ and $y_\mu$, are real and positive, but nontrivial phases can still be present in $y^\pm_{LE,EL}$. The fact that four complex phases in Eq.~(\ref{eq:fermion-yukawa}) are physical can be seen through parameter counting: there are 13 Yukawa couplings and 10 fermion fields that may undergo chiral phase rotations without affecting any other parameters in the action. Since a uniform rotation of all fields leaves Eq.~(\ref{eq:fermion-yukawa}) invariant, 9 complex phases can be eliminated, leaving the four physical phases in $y^\pm_{LE,EL}$.

After spontaneous symmetry breaking, 
%we can parameterize 
the scalar fields $S$ and $H$ in the unitary gauge are $S = (v_S + h_D)/\sqrt{2}$ and $H=(0,v + h)/\sqrt{2}$, %\RX{For consistency, either the non-zero VEV component of the Higgs is on the top component here, or we have to flip the components of the lepton field in table I.}
where $v_S$ and $v$ are the vev's of the dark and SM Higgs fields, respectively. 
%, and $h_D$ and $h$ are their remaining physical degrees of freedom. 
Writing the fermion fields as $\psi^\mu_{L,R} = (\mu_{L,R}, \, D^+_{L,R},\, E^+_{L,R},\, D^-_{L,R},\, E^-_{L,R})^T$, $\psi^\nu_L = (\nu_L, \nu^+_L, \nu^-_L)$, and $\psi^\nu_R = (\nu^+_R, \nu^-_R)$, Eq.~(\ref{eq:fermion-yukawa}) yields mass terms of the form $\overline{\psi}^\mu_L \mathcal{M}^\mu \psi^\mu_R$ and $ \overline{\psi}^\nu_L \mathcal{M}^\nu \psi^\nu_R$, with
\begin{align}\label{eq:mass-mats}
  %  &-\overline{\psi}^\mu_L \mathcal{M}^\mu \psi^\mu_R - \overline{\psi}^\nu_L \mathcal{M}^\nu \psi^\nu_R + h.c.,\\
    &\mathcal{M}^\mu \equiv \frac{1}{\sqrt{2}}\begin{pmatrix}
    y_\mu v & y^+_L v_S & 0 & y^-_L v_S & 0\\
    0 & \sqrt{2}M^+_L & y^+_{LE} v & 0 & 0\\
    y^+_E v_S & y^+_{EL} v & \sqrt{2} M^+_E & 0 & 0\\
    0 & 0 & 0 & \sqrt{2} M^-_L & y^-_{LE} v\\
    y^-_E v_S & 0 & 0 & y^-_{EL} v & \sqrt{2} M^-_E
    \end{pmatrix}, \nonumber\\
    &\mathcal{M}^\nu \equiv \frac{1}{\sqrt{2}} \begin{pmatrix}
    y^+_L v_S & y^-_L v_S\\
    \sqrt{2} M^+_L & 0\\
    0 & \sqrt{2} M^-_L
    \end{pmatrix}.
\end{align}
% \begin{align}\label{eq:mass-mats}
%   %  &-\overline{\psi}^\mu_L \mathcal{M}^\mu \psi^\mu_R - \overline{\psi}^\nu_L \mathcal{M}^\nu \psi^\nu_R + h.c.,\\
%     &\mathcal{M}^\mu \equiv \begin{pmatrix}
%     \frac{y_\mu v}{\sqrt{2}} & \frac{y^+_L v_S}{\sqrt{2}} & 0 & \frac{y^-_L v_S}{\sqrt{2}} & 0\\
%     0 & M^+_L & \frac{y^+_{LE} v}{\sqrt{2}} & 0 & 0\\
%     \frac{y^+_E v_S}{\sqrt{2}} & \frac{y^+_{EL} v}{\sqrt{2}} & M^+_E & 0 & 0\\
%     0 & 0 & 0 & M^-_L & \frac{y^-_{LE} v}{\sqrt{2}}\\
%     \frac{y^-_E v_S}{\sqrt{2}} & 0 & 0 & \frac{y^-_{EL} v}{\sqrt{2}} & M^-_E
%     \end{pmatrix},\\
%     &\mathcal{M}^\nu \equiv \begin{pmatrix}
%     \frac{y^+_L v_S}{\sqrt{2}} & \frac{y^-_L v_S}{\sqrt{2}}\\
%     M^+_L & 0\\
%     0 & M^-_L
%     \end{pmatrix}. \nonumber
% \end{align}
%
%Bidiagonalizing these mass matrices in the usual manner with unitary matrices $U^{\mu,\nu}_{L,R}$, we have
Bidiagonalizing these matrices in the usual manner, we have
\begin{align}
    (U^{\mu})^\dagger_L \mathcal{M}^{\mu} U^{\mu}_R = \textrm{diag}(m_\mu, m^+_D, m^+_E, m^-_D, m^-_E).
    \\
    (U^{\nu})^\dagger_L \mathcal{M}^{\nu} U^{\nu}_R = \begin{pmatrix}
   0 & 0\\
    m^+_\nu & 0\\
    0 & m^-_\nu
    \end{pmatrix}, 
\nonumber
\end{align}
where the $m^\pm_\nu$ are of the order $m^\pm_D$, which further implies that the new vectorlike doublets do not contribute appreciably to the electroweak $T$ parameter.

As seen shortly, experimental searches for leptonic portal matter generally require $M^\pm_{L,E} \gsim 1 \; \textrm{TeV}$. Assuming that $m_{A_D} = g_D v_S \sim O(0.01-1 \; \textrm{GeV})$, this in turn suggests that $v_S/M^\pm_{L,E} \sim O(10^{-(3-4)})$. Meanwhile, $y_\mu v/M^\pm_{L,E}$ and $y^\pm_{LE, EL} v/M^\pm_{L,E}$ may in principle be somewhat large if these Yukawa couplings saturate perturbativity bounds (in which case $y v/\sqrt{2} \sim 600 \; \textrm{GeV}$), but in practice the parameter space of interest has $y_\mu, \vert y^\pm_{LE,EL}\vert \ll 1$. Hence, we can sensibly perform our analysis perturbatively in the limit that $y_\mu v,\; \vert y^\pm_{LE,EL}\vert v,\; y^\pm_{L,E}v_S \ll M^\pm_{L,E}$. To leading order, we then have $ m_\mu \approx \frac{y_\mu v}{\sqrt{2}}$, $m^\pm_D \approx M^\pm_L$, $m^\pm_E \approx M^\pm_E$,$m^\pm_\nu \approx M^\pm_L$,
%%
%\begin{align}
%    m_\mu \approx \frac{y_\mu v}{\sqrt{2}}, \;\;\; m^\pm_D \approx M^\pm_L, \;\;\; m^\pm_E \approx M^\pm_E, \;\;\; m^\pm_\nu \approx M^\pm_L,
%\end{align}
%
%with each of the above results having 
up to proportional corrections of $O(v^2/(M^{\pm}_{L,E})^2,v_S^2/(M^{\pm}_{L,E})^2,v_S v/(M^{\pm}_{L,E})^2)$. The coupling of the portal matter fields to SM matter is described in detail in \cite{Rizzo:2018vlb}, so we shall only summarize the critical points here: The SM isospin doublet (singlet) fermion achieves $O(v_S/M^{\pm}_{L,E})$ mixing with the isospin doublet (singlet) portal matter fields and, in the case of the charged leptons, $O(m_\mu v_S/(M^{\pm}_{L,E})^2)$ mixing with the isospin singlet (doublet) portal matter. These mixings facilitate nearly, but not entirely, chiral couplings between portal matter and SM fermions mediated by the dark photon, dark Higgs, and heavy electroweak bosons. As the dark Higgs possesses an $O(1)$ coupling between the portal matter and the SM, the portal matter overwhelmingly decays to SM fields via the emission of a dark Higgs or the longitudinal mode of a dark photon, leading to a distinctive collider signature of a high-$p_T$ lepton and, depending on the decay of the dark photon and dark Higgs, either missing energy or a pair of highly collinear electrons, muons, or pions.
\section{\label{sec:g-2-calc} $g-2$ Calculation}
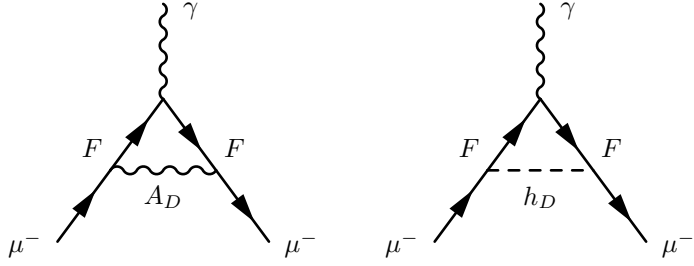
\begin{figure*}[ht!]
\centering
\subfloat{
%\LARGE
\begin{fmffile}{feyngraph}
  \begin{fmfgraph*}(80,100)
    \fmfbottom{f2,f1}
    \fmftop{g}
    \fmf{boson,t=1.5}{g,v}
    \fmf{fermion}{f2,v1}
    \fmf{fermion}{v1,v}
    \fmf{fermion} {v,v2}
    \fmf{fermion}{v2,f1}
    \fmfv{l.a=-20,l.d=8,l=$\gamma$}{g}
    \fmfv{l.a=-10,l=$\mu^-$}{f1}
    \fmfv{l.a=-170,l=$\mu^-$}{f2}
    %\fmflabel{$\mu^-$}{f1}
    %\fmflabel{$\mu^-$}{f2}
    \fmfv{l.a=120, d=10,l=$F$}{v1}
    \fmfv{l.a=60, d=10,l=$F$}{v2}
    \fmffreeze
    \fmf{boson,label=$A_D$}{v1,v2}
  \end{fmfgraph*}
  \end{fmffile}
}\hspace{15mm}
\subfloat{
%\LARGE
\begin{fmffile}{feyngraph2}
  \begin{fmfgraph*}(80,100)
    \fmfbottom{f2,f1}
    \fmftop{g}
    \fmf{boson,t=1.5}{g,v}
    \fmf{fermion}{f2,v1}
    \fmf{fermion}{v1,v}
    \fmf{fermion} {v,v2}
    \fmf{fermion}{v2,f1}
    \fmfv{l.a=-20,l.d=8,l=$\gamma$}{g}
    \fmfv{l.a=-10,l=$\mu^-$}{f1}
    \fmfv{l.a=-170,l=$\mu^-$}{f2}
    \fmfv{l.a=120, d=10,l=$F$}{v1}
    \fmfv{l.a=60, d=10,l=$F$}{v2}
    \fmffreeze
    \fmf{dashes,label=$h_D$}{v1,v2}
  \end{fmfgraph*}
  \end{fmffile}
}
\caption{The Feynman diagrams contributing to the muon $g-2$ corrections, with $F=\{D^\pm,E^\pm \}$.}\label{fig1}
\end{figure*}

%
%With the model in place, we can now discuss the computation of the muon $g-2$. 
In this model, there are several diagrams which yield new physics contributions at one loop to $g-2$. However, as seen in \cite{Rizzo:2018vlb}, the ``conventional'' dark photon contribution, arising from a dark photon emitted from an internal muon line in the loop (see, for example, \cite{Pospelov:2008zw,Davoudiasl:2014kua}) is generally between one and two orders of magnitude too small to account for the muon anomaly. In our framework, for $m_{A_D} \leq 1 \; \textrm{GeV}$, this contribution can account for $\lesssim 10 \%$ of the observed anomalous muon magnetic moment and remain consistent with constraints on kinetic mixing from beam dump experiments, for example NA64 \cite{NA64:2021xzo}. Instead, the dominant contributions to the magnetic moment must come from elsewhere -- specifically, loops with a heavy vector-like fermion in the internal line. Diagrams of this class featuring the heavy electroweak bosons, the dark Higgs, and the dark photon can all contribute. However, in practice the overwhelmingly dominant contributions  emerge from the dark Higgs and the longitudinal polarization of the dark photon, for the same reason that emissions of these particles dominate the decay width of the portal matter:  because the dark Higgs possesses an $O(1)$ coupling between portal matter and SM states, these interactions (and by Goldstone boson equivalence, those with the longitudinal mode of the dark photon)  provide the strongest couplings of the portal matter to the SM. As $m_\mu, m_{A_D}, m_{h_D} \ll M^\pm_{L,E}$, the general results for $a_\mu$  \cite{Leveille:1977rc} in this limit yield extremely simple expressions. More precisely, for a given heavy fermion $F$ which couples to the muon via
\begin{align}
    &g_L A_D^\mu \overline{\mu}_L \gamma_\mu F_L + g_R A_D^\mu \overline{\mu}_R \gamma_\mu F_R \nonumber \\ 
    &+ y_{LR} h_D \overline{\mu}_L F_R + y_{RL} h_D \overline{\mu}_R F_L + h.c., 
\end{align}
the contribution of Feynman diagrams with a dark photon loop and a dark Higgs loop with $F$ on the internal fermion lines, as shown in Fig.~\ref{fig1}, is
%
%\begin{widetext}
\begin{align}\label{eq:approx-g}
    \Delta a^{A_D}_\mu &\approx \frac{m_\mu^2}{16 \pi^2} \bigg( -\frac{5}{6} \frac{(|g_L|^2+|g_R|^2)}{m_{A_D}^2} + \frac{m_F}{m_\mu} \frac{Re[g_L^* g_R]}{m_{A_D}^2}\bigg), \nonumber \\
    \Delta a^{h_D}_\mu &\approx \frac{m_\mu^2}{16 \pi^2} \bigg( \frac{(|y_{LR}|^2+|y_{RL}|^2)}{6 m_F^2} + \frac{m_F}{m_\mu} \frac{Re[y_{LR}^* y_{RL}]}{m_F^2}\bigg).
\end{align}
%\end{widetext}
%

The Yukawa couplings $y_{LR,RL}$ in Eq.~(\ref{eq:approx-g}) are governed by $y^\pm_{LE,EL}$, which as we recall can be complex.
%Up to now, we have (without loss of generality) taken a basis in which all Yukawa couplings and mass parameters are real and positive, with the exception of $y^\pm_{LE}$ and $y^\pm_{EL}$, which can be complex.
However, phases in $y^\pm_{LE,EL}$ will generally also 
%lead to the same dark photon and dark Higgs loops 
contribute to the \emph{electric} dipole moment of the muon, which will in turn contribute to the measured anomalous magnetic dipole moment \cite{Feng:2001sq}. In the absence of significant tuning this contribution will be highly subleading. While the bounds on the muon EDM will be improved at the J-PARC E34 experiment \cite{Abe:2019thb} as well as at the FNAL Muon g-2 experiment, a direct measurement of this effect might only be reached
in a dedicated muon EDM experiment, such as the planned experiment at PSI \cite{Adelmann:2021udj}. 
% \cite{Farley:2003wt}. 

Hence, in this what follows we shall assume for simplicity that $y^\pm_{LE}$ and $y^\pm_{EL}$ are real, but may have arbitrary signs-- for practical purposes the only difference between our results here and the more general case with complex Yukawa couplings (absent a direct measurement of muon EDM) would be that the dominant contributions of the new physics to the $\Delta a_\mu$ would be scaled by cosines of the phase differences between these Yukawa couplings.

Of particular importance are the terms in Eq.~(\ref{eq:approx-g}) bearing the chiral enhancement factor $m_F/m_\mu \gg 1$. Without any information about the coupling constants, we would expect these terms to overwhelmingly dominate the $g-2$ correction. However, for isospin doublet (singlet) portal matter, $g_L (g_R)$ and $y_{RL} (y_{LR})$ are suppressed relative to their opposite-chirality couplings by a factor of $O(v/M^{\pm}_{L,E})$, so the chirally enhanced terms contribute to $g-2$ at the same order in $v/M^{\pm}_{L,E},v_S/M^{\pm}_{L,E}$ as those which do not bear a chiral enhancement. This behavior of chirality-flipping operators in models with vector-like leptons is, of course, well-explored; it is precisely analogous to the behavior noted in \cite{Dermisek:2013gta,CarcamoHernandez:2019ydc,Crivellin:2021rbq,Crivellin:2018qmi}. Computing the relevant dark photon and dark Higgs couplings in our model's mass basis by bidiagonalizing the mass matrices of Eq.~(\ref{eq:mass-mats}), we find that
\begin{align}\label{eq:delta-a-model-result-parts}
    \Delta a^{A_D,D^\pm}_\mu = \Delta a^{h_D,D^\pm}_\mu = \frac{m_\mu^2 (y^\pm_L)^2}{192 \pi^2 (M^\pm_L)^2} \bigg( 1 + \eta^\pm_L \bigg),\\
    \eta^\pm_L \equiv \frac{6}{y_\mu}\frac{y^\pm_E}{y^\pm_L} \frac{M^\pm_L (M^\pm_E y^\pm_{LE} + M^\pm_L y^\pm_{EL})}{(M^\pm_L)^2-(M^\pm_E)^2} \nonumber,
\end{align}
where the contributions from the weak isospin singlet portal matter, $\Delta a^{A_D,E^\pm}_\mu$ and $\Delta a^{h_D,E^\pm}_\mu$, are given by analogous expressions to Eq.~(\ref{eq:delta-a-model-result-parts}), but with $L \longleftrightarrow E$ (with the caveat that $y^{\pm}_{LE}$ and $y^{\pm}_{EL}$ are not interchanged, and noting that $y^\pm_{LE}$ and $y^\pm_{EL}$ are real and can be either positive or negative). %, and recalling that $y^\pm_{LE}$ and $y^\pm_{EL}$ are real and can have either sign. 
In these expressions, we have used $m_{A_D} = g_D v_S$ and $m_\mu \approx y_\mu v/\sqrt{2}$. We further note that although the individual expressions for $\Delta a^{A_D, D^\pm}$ and $\Delta a^{h_D, D^\pm}$ in Eq.~(\ref{eq:delta-a-model-result-parts}) may appear to diverge in the limit that $M^\pm_L \rightarrow M^\pm_E$, these apparent divergences are always cancelled in the full expression for $\Delta a_\mu$ by the corresponding terms in $\Delta a^{A_D, E^\pm}$ and $\Delta a^{h_D, E^\pm}$.

The equivalence between the dark photon and dark Higgs contributions to $g-2$ is unsurprising. In the limit that we have taken in the unitary gauge (specifically that $m_{h_D}, m_{A_D} \ll M^\pm_{L,E}$), the only contributing part of the dark photon propagator in the loop stems from the longitudinal polarization, which in the Feynman gauge is simply the Goldstone boson part of the scalar $S$, of which the dark Higgs $h_D$ forms the physical part. 

More notably, the fact that the new physics $g-2$ contribution in this model is entirely dominated by the dark Higgs (and its Goldstone boson in the form of the longitudinal mode of the dark photon) means that the $g-2$ result in this model is to leading order \emph{independent} of any of the parameters which govern the relic abundance or detection phenomenology of the dark sector: The dark gauge coupling, kinetic mixing, and even the masses of the dark photon, dark Higgs, and dark matter fields do not enter into Eq.~(\ref{eq:delta-a-model-result-parts}), as long as they remain within the approximate order of magnitude as specified by our parameter space. 
This conclusion even remains if one considers more exotic methods of generating the relic abundance than freeze-out dominated by simple $s$-channel annihilation to SM fermions, for example freeze-out dominated by kinematically forbidden annihilations or $3 \rightarrow 2$ processes \cite{Fitzpatrick:2020vba,Wojcik:2021xki,Hara:2021lrj,DAgnolo:2015ujb} -- Eq.~(\ref{eq:delta-a-model-result-parts}) \emph{only} depends on the portal matter masses and their mixings with the SM fermions, and not on any of the parameters of the simplified dark matter model. This insensitivity to the specifics of the dark sector is in marked contrast to models that recreate the muon $g-2$ anomaly using lepton portal dark matter such as that of \cite{Kawamura:2022uft}.

Combining all of these contributions, we arrive at the total new physics muon $g-2$ contribution in the model (up to sub-leading corrections from finite dark photon and dark Higgs masses and numerically subdominant diagrams, such as loops with the heavy electroweak bosons), which is given by
\begin{align}\label{eq:delta-a-model-result}
    \Delta a^{\rm PM}_\mu &= \Delta a^+_\mu + \Delta a^-_\mu,\\ 
    \Delta a^\pm_\mu &\equiv \frac{m_\mu^2}{96 \pi^2} \bigg( \frac{(y^\pm_L)^2}{(M^\pm_L)^2} + \frac{(y^\pm_E)^2}{(M^\pm_E)^2} - \frac{6 y^\pm_L y^\pm_E}{M^\pm_L M^\pm_E} \frac{y^\pm_{LE}}{y_\mu} \bigg), \nonumber
\end{align}
where for clarity we have separated the contributions from the $+1$-charged portal matter, $a^+_\mu$ from those of the $-1$-charged portal matter, $a^-_\mu$. Note that while the contributions to $g-2$ from the Yukawa couplings $y^\pm_{EL}$ cancel in the summation, those corresponding to the SM Higgs Yukawa couplings with the same chiral structure as $y_\mu$, namely $y^\pm_{LE}$, persist in the final expression for $\Delta a_\mu$. Using Eq.~(\ref{eq:delta-a-model-result}), we can now perform some numerical estimates of the magnitude of the effect. As we previously discussed, collider searches for leptonic portal matter yield strong constraints, requiring $M^\pm_{L,E} \gsim 1 \; \textrm{TeV}$. Since perturbativity in turn requires that $y^\pm_{L,E}$ must be no greater than $O(1)$, the first two terms in the expression for $\Delta a^\pm_\mu$ in Eq.~(\ref{eq:delta-a-model-result}) can only feasibly contribute at roughly the same level as the dark photon loop with an internal muon, that is to say, they might account for $1-10\%$ of the anomalous magnetic moment. Isolating the third term, we have
\begin{align}\label{eq:delta-a-result-num}
    \Delta a^\pm_\mu \approx -(\Delta a_\mu)
    %(2.5 \times 10^{-9}) 
    \bigg(\frac{y^\pm_{LE}/y_\mu}{36}\bigg) \bigg(\frac{1 \; \textrm{TeV}}{M^\pm_E/y^\pm_E} \bigg)\bigg(\frac{1 \; \textrm{TeV}}{M^\pm_L/y^\pm_L} \bigg),
\end{align}
where $\Delta a_\mu$ is the measured anomaly described in Eq.~(\ref{eq:delta-a}). Hence, the observed $g-2$ anomaly can be accommodated as long as at least one of the Yukawa couplings $y^+_{LE}$ or $y^-_{LE}$ are negative (recall that we are working in a basis in which all Yukawa couplings are real and positive except $y^\pm_{LE}$ and $y^\pm_{EL}$),  %but in general those Yukawa couplings may be negative and even have a complex phase, with the caveat that any imaginary part to these couplings can potentially contribute to the muon electric dipole moment), 
and there is a modest $O(10)$ hierarchy between whichever coupling(s) $y^\pm_{LE}$ dominate the $g-2$ contribution and $y_\mu$. As $y_\mu$ is very small (roughly $O(10^{-4})$), allowing $y^\pm_{LE}$ to achieve a larger value is trivial -- for $M^\pm_{L,E} \sim 1 \; \textrm{TeV}$, we expect  
%the mass term arising from $y^\pm_{LE}$ to be
$\vert y^\pm_{LE}\vert v \sim 
O(\textrm{a few} \;\! \textrm{GeV})$, 
in contrast to the $O(m_t)$ values for this quantity needed in \cite{CarcamoHernandez:2019ydc}, and roughly on par with the SM Yukawa couplings of the $c$ and $b$ quarks, or the $\tau$ lepton. 
%As $v \sim 246 \; \textrm{GeV}$, 
These values are easily achievable with perturbative $y^\pm_{LE}$ values. Since these Yukawa couplings can only appear if there is a complete vector-like lepton family (with both an isospin doublet and a singlet) mixing with the muon, it is also clear why the minimal construction of \cite{Rizzo:2018vlb} could not reproduce the observed $g-2$ anomaly.

As mentioned in Section \ref{sec:intro}, the method by which we have achieved a sizable $g-2$ correction here bears a significant similarity to models which address the muon $g-2$ anomaly with heavier $O(100-1000 \; \textrm{GeV})$ fermiophobic $Z'$ bosons, in particular \cite{CarcamoHernandez:2019ydc}. There the authors use the same chiral enhancement mechanism to generate the $g-2$ anomaly. However, our choice of a light dark photon inspired by sub-GeV kinetic mixing/vector portal dark matter, instead of a heavy hidden $Z'$, leads to some unique characteristics: First, the contributions to the $g-2$ in Eq.~(\ref{eq:delta-a-model-result}) are entirely independent of the gauge sector parameters such as the dark photon mass and its associated gauge coupling, ultimately due to a version of Goldstone boson equivalence. %-- while it is somewhat obfuscated in the unitary gauge, a repeat of this calculation in the Feynman gauge makes clear that the dominant contribution to $g-2$ is entirely determined by lepton mass matrix parameters. 
Second, and perhaps more significantly, we shall see that the phenomenological constraints on this model differ markedly from the construction in \cite{CarcamoHernandez:2019ydc} and similar works.

\section{Other Constraints}

It is clear from our result for $\Delta a_\mu$ in Eq.~(\ref{eq:delta-a-model-result}) that the principal parameters that govern $\Delta a_\mu$ are the Yukawa couplings $y^{\pm}_{LE}$ and the masses of the vector-like leptons, approximately given by $M^{\pm}_{L,E}$. Notably, the simplified sub-GeV dark matter model
%, which must reproduce the observed dark matter relic abundance, 
has no dependence on any of the parameters in Eq.~(\ref{eq:delta-a-model-result}), except for a logarithmic dependence of the kinetic mixing parameter $\epsilon$ on the ratios $M^+_{L,E}/M^-_{L,E}$. It is therefore trivial to find specific simplified dark matter models which reproduce the observed relic abundance and are accommodated by this construction.
%, meaning that the dark sector provides no meaningful constraint on the model.

\begin{figure}
    \centering
    \includegraphics[width=8cm]{"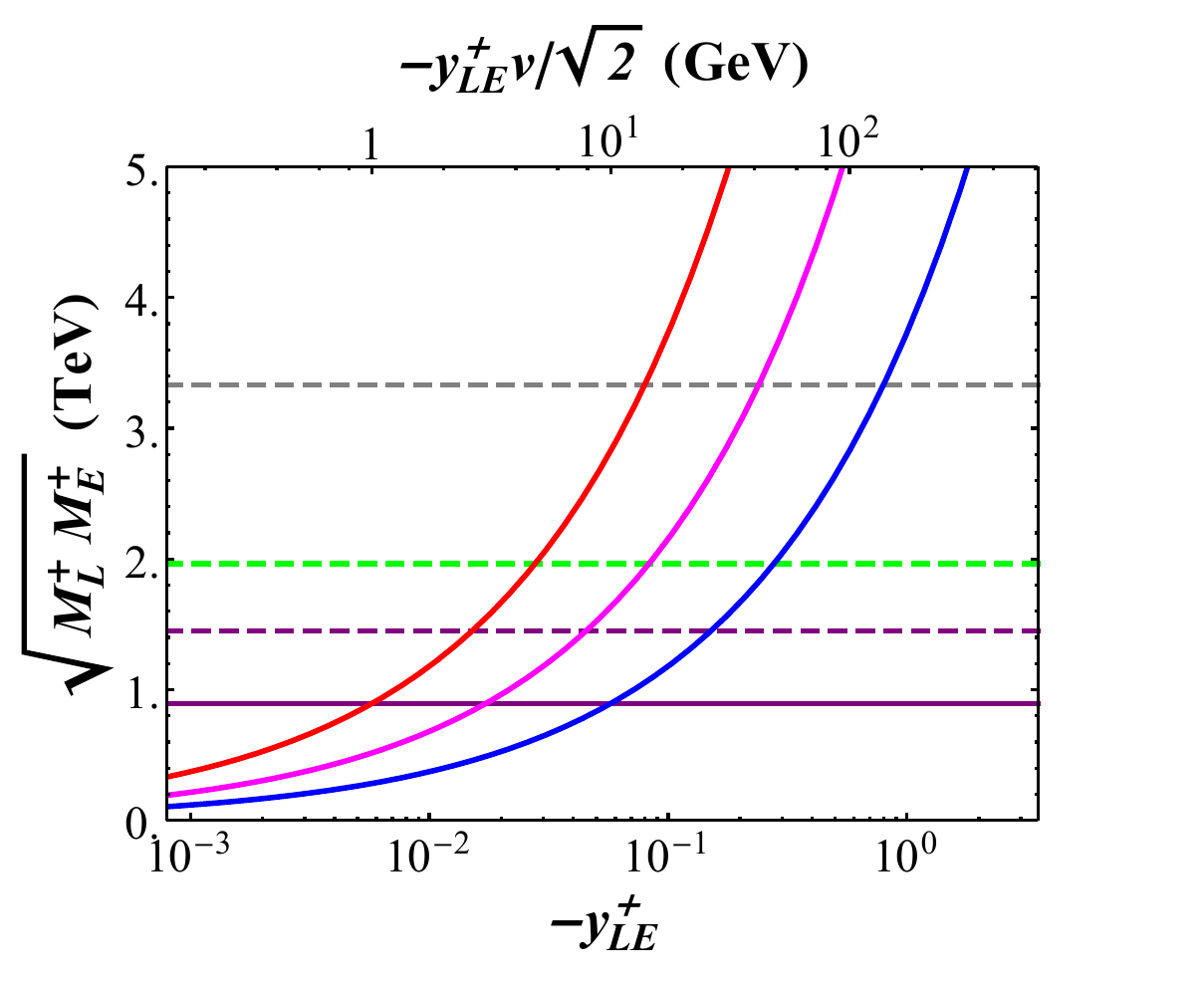"}
    \caption{\footnotesize The mass scale of the $+1$-charged vector-like leptons ($\sqrt{M^+_L M^+_E}$) to produce $\Delta a_\mu = 2.51 \times 10^{-9}$, as a function of $y^+_{LE}$ (assuming $\vert y^+_{LE}\vert \gg \vert y^-_{LE}\vert$), for $y^+_L y^+_E = 0.3$ (blue), $1$ (magenta) and $3$ (red). 
    %We have taken $y^+_{LE} \gg y^-_{LE}$, so that $-1$-charged portal matter states do not contribute sizably to $\Delta a_\mu$. 
    The minimum LHC mass of $M^+_E$ from \cite{Guedes:2021oqx} is depicted by a solid purple line, while the projected limits from \cite{Guedes:2021oqx} on $M^+_E$ are dashed lines for the HL-LHC (purple), HE-LHC (green), and $hh$-FCC (gray).}
    \label{fig2}
\end{figure}

Instead of the dark matter sector, we must turn to the portal matter sector directly for experimental constraints. Inspecting Eq.~(\ref{eq:delta-a-result-num}), we see that we may constrain this model by considering limits on the vector-like lepton masses. As previously stated, the current constraints on vector-like leptonic portal matter are already much stronger than the analogous limits on vector-like leptons which dominantly decay into heavier electroweak bosons, because the portal matter will overwhelmingly decay via the emission of light dark photons and dark Higgses. Assuming that these fields either decay into dark matter or are long-lived enough to escape a detector, the experimental signature for portal matter production strongly resembles that of sleptons. The study in \cite{Guedes:2021oqx} recasts the null results of an ATLAS slepton search with $139\; \textrm{fb}^{-1}$ of data at $13 \; \textrm{TeV}$, \cite{ATLAS:2018ojr}, as a limit on the mass of isospin singlet vector-like leptons, finding in our notation a limit of $M^\pm_E \geq 895 \; \textrm{GeV}$, with the mass constraints for vector-like leptons being about $16-20\%$ higher due to larger production cross sections \cite{Rizzo:2022qan,OsmanAcar:2021plv}. Assuming that $y^\pm_{L,E} \sim O(1)$, this constraint informs our estimate that $\vert y^\pm_{LE}\vert v \sim O( \textrm{GeV})$ to achieve the observed value of $\Delta a_\mu$. While a detailed study is beyond the scope of this work, we can get a feel for the parameter space by performing some rough estimates. Fig.~\ref{fig2} shows the required mass scale of the $+1$-charged portal matter, $\sqrt{M^+_L M^+_E}$, to reproduce $\Delta a_\mu = 2.51 \times 10^{-9}$, as a function of the Yukawa coupling $y^+_{LE}$, assuming that $\vert y^+_{LE} \vert \gg \vert y^-_{LE}\vert $, so that the contribution of the $+1$-charged portal matter fields to $\Delta a_\mu$ dominates that of the $-1$-charged portal matter. We have also included current and projected hadron collider constraints on isospin singlet leptonic portal matter from \cite{Guedes:2021oqx}.

Notably, we see that even relatively near-term experiments, such as the HL-LHC, can probe many modest values of $\vert y^\pm_{LE}\vert v \sim O(\textrm{GeV})$. Longer term hadron collider experiments, such as $hh$-FCC, might be able to probe $y^\pm_{LE}$ values comparable to the top quark Yukawa coupling. We also note that a multi-TeV muon collider, such as discussed in \cite{Black:2022cth}, might permit probes of even higher portal matter lepton masses, up to roughly half of that collider's center-of-mass energy \cite{OsmanAcar:2021plv}. As such, even a modest $2 \; \textrm{TeV}$ muon collider provides comparable $y^\pm_{LE}$ reach to the current LHC constraints, while a muon collider with a center-of-mass energy $\gsim 6 \; \textrm{TeV}$ will generally provide better reach than even the $hh$-FCC.

Several constraints that are relevant to other models that generate $\Delta a_\mu$ with vector-like leptons are highly suppressed in this model. For example, as the coupling between the SM muon-neutrino and the dark photon is suppressed by $v_S^2/(M^\pm_{L,E})^2 \lesssim 10^{-(6-8)}$, significant constraints such as neutrino trident production \cite{Altmannshofer:2014pba} and bounds on muon-neutrino scattering from Borexino \cite{Bellini:2011rx} are several orders of magnitude too weak to significantly constrain this model, in marked contrast to the case of 
%, \eg, the neutrino trident constraints in 
\cite{CarcamoHernandez:2019ydc}. In addition, modifications to %to Higgs processes such as 
$h \rightarrow \gamma \gamma$ and $h \rightarrow \mu^+ \mu^-$ are suppressed by $v_S^2/(M^\pm_{L,E})^2 \lesssim 10^{-(6-8)}$ and $v_S v/(M^\pm_{L,E})^2 \lesssim 10^{-(4-5)}$, respectively, which again are too weak to be presently constrained, though they are significant in the scenarios of \cite{Dermisek:2013gta}.

\section{Summary and Conclusions}
In this letter, we have discussed a simple model of portal matter that can account for the observed muon $g-2$ anomaly via a chiral enhancement factor similar to that of \cite{CarcamoHernandez:2019ydc}. However, in our model the sub-GeV mass scale of the dark photon, motivated by the kinetic mixing/vector portal dark matter scenario, has a profound influence on the $g-2$ contribution and the other phenomenological constraints. Most notably, due to the light dark photon mass in this theory, $\Delta a_\mu$ is to leading order independent of \emph{any} parameters which appear in the simplified dark matter model, including the kinetic mixing, the masses of the dark photon and dark Higgs, and even the dark gauge coupling. As in \cite{CarcamoHernandez:2019ydc}, the contribution to $\Delta a_\mu$ is dominated by a SM Higgs Yukawa coupling between the isospin singlet and doublet portal matter fields, $y^\pm_{LE}$. However, the favored parameter range for $y^\pm_{LE}$ in this model is remarkably small, roughly comparable to the $\tau$ lepton Yukawa coupling, rather than $O(1)$. As $\Delta a_\mu$ depends on only a few parameters ($y^\pm_{LE}$ and the portal mass parameters $M^\pm_{L,E}$), large portions of the parameter space of this model can be readily explored with searches for leptonic portal matter fields at either hadron or lepton colliders; multi-TeV muon colliders in particular represent a promising probe.

Our model is a minimal portal matter construction that addresses the muon $g-2$ anomaly, and it would be interesting to explore further extensions. A more detailed study within this framework might consider the constraints arising from mixing of the portal matter with the electron, as in \cite{CarcamoHernandez:2019ydc}, or how one might naturally satisfy the constraints in a more elaborate construction. Meanwhile, the similarity between our favored values for the Yukawa coupling $y^\pm_{LE}$ (for portal matter masses of $\sim 1 \; \textrm{TeV}$) and the Yukawa coupling for the $\tau$ lepton are suggestive that the order of magnitude of of $y^\pm_{LE}$ might be the consequence of some sort of flavor symmetry \cite{Wojcik:2022xxx}.

\section*{Acknowledgements}
This work was supported by the U.S. Department of Energy under the contract number DE-SC0017647.

\bibliography{main}% Produces the bibliography via BibTeX.

\end{document}